\documentclass[12pt]{article}
\usepackage{graphicx}
\begin{document}
\baselineskip=5.5mm
\newcommand{\be} {\begin{equation}}
\newcommand{\ee} {\end{equation}}
\newcommand{\Be} {\begin{eqnarray}}
\newcommand{\Ee} {\end{eqnarray}}
\renewcommand{\thefootnote}{\fnsymbol{footnote}}
\def\a{\alpha}
\def\b{\beta}
\def\g{\gamma}
\def\G{\Gamma}
\def\d{\delta}
\def\D{\Delta}
\def\e{\epsilon}
\def\k{\kappa}
\def\l{\lambda}
\def\L{\Lambda}
\def\t{\tau}
\def\om{\omega}
\def\Om{\Omega}
\def\s{\sigma}
\def\lg{\langle}
\def\rg{\rangle}
\noindent
\begin{center}
{\Large
{\bf
Fluctuation-dissipation relations for a general class of master equations
}}\\
\vspace{0.5cm}
\noindent
{\bf Gregor Diezemann} \\
{\it
Institut f\"ur Physikalische Chemie, Universit\"at Mainz,
Welderweg 11, 55099 Mainz, FRG\\}
\end{center}

\vspace{1cm}
\noindent
{\it
The fluctuation-dissipation relation is calculated for a class of stochastic
models obeying a master equation. The transition rates are assumed to obey
detailed balance also in the presence of a field.
It is shown that in general the linear response cannot be expressed via
time-derivatives of the correlation function alone, but an additional function
$\xi(t,t_w)$, which has been rarely discussed before is required.
This function depends on the two times also relevant for the response and the
correlation and vanishes under equilibrium conditions.
It can be expressed in terms of the propagators and the transition rates of the
master equation but it is not related to any physical observable in an obvious
way.
Instead, it is determined by inhomogeneities in the temporal evolution of the
distribution function of the stochastic variable under consideration.
$\xi(t,t_w)$ is considered for some examples of stochastic models, some of
which exhibit true non-equilibrium dynamics and others approach equilibrium
in the long term. In particular, models in which a relevant variable, e.g. a
magnetization, is related in a prescribed way to the states of the system are
considered as projections from a composite Markov process.
From these model calculations, it is conjectured that $\xi(t,t_w)$ vanishes
when one is concerned with measurements of the analogue of a structure factor
for large wave-vectors or if every transition among the states randomizes the
value of the stochastic variable considered.
}

\vspace{0.5cm}
\noindent
PACS Numbers: 64.70 Pf,05.40.+j,61.20.Lc
\vspace{1cm}
\section*{I. Introduction}
The out-of-equilibrium dynamics of stochastic models has gained intensive
interest in the last decade. In particular, the deviations from the
fluctuation dissipation theorem (FDT), relating the linear response to the
two-time correlation function have been investigated in great detail, for a
recent review see\cite{CR03}.
Starting with an investigation of the spherical $p$-spin-glas model\cite{CK93},
much attention has been paid to study the behavior of the response and the
correlation for models of glassy dynamics.
While in equilibrium the FDT relates the response to the correlation in a
unique way, this does not hold in out-of-equilibrium situations.
The violations of the FDT usually are parameterized via the introduction of
a function $X(t,t_w)$, which is defined via:
\be\label{X_FDT}
R(t,t_w)={X(t,t_w)\over T}{\partial C(t,t_w)\over\partial t_w}
\ee
Here, the correlation function of a quantity $M(t)$ is defined by
$C(t,t_w)\!=\!\lg M(t)M(t_w)\rg$ and the corresponding response to a field
conjugate to $M(t)$ is $R(t,t_w)\!=\!\d\lg M(t)\rg/\d H(t_w)|_{H=0}$
for $t\!\geq\!t_w$.
In case that $M$ is a so called neutral variable\cite{FS02}, $X(t,t_w)$ is
independent of $M$ and the long time limit $X_\infty$ allows the definition
of an effective temperature\cite{CKP97}.

The value of $X_\infty$ has been calculated for a variety of models with
different results.
One class of models that have been considered are coarsening
models\cite{bray94}, for which $X_\infty$ is known to vanish\cite{X_infty}.
Examples of such models are the well known spherical model\cite{BK52,GL00}
and the $O(N)$ model in the limit of large $N$\cite{CLZ02}, as well as the
Ising models in one\cite{GL00Is} or higher dimension\cite{IsRev}.
In the context of models for glassy dynamics for some discontinuous mean field
spin models a different behavior has been found\cite{FDT_rev98}.
In particular, a relation between the degree of replica symmetry breaking
and $X_\infty$ has been established.
In addition, in some time sectors, in which both, the response and the
correlation obey some scaling relations, it has been found that $X(t,t_w)$
is a function of the correlation alone, $X(C)$.
For disordered systems, a connection between the out-of-equilibrium dynamics
and static properties has been established\cite{FMPP98}.
For coarsening systems, however, such a relation does not exist\cite{CLZ01}.
In addition to these analytical calculations, a number of molecular dynamics
simulations have been performed on model glassforming liquids, for a recent
review see ref.\cite{barrat03}.

Some of the quoted models are soft-spin models in the sense that the stochastic
dynamics is calculated from a Langevin equation. The classical treatment of
FDT violations for unfrustrated stochastic models with Langevin dynamics has
been given in ref.\cite{CKP94}, where various examples, including a simple
random walk have been considered.
In addition to models obeying a Langevin equation, the out-of-equilibrium
dynamics of models with a dynamics determined by a master
equation\cite{vkamp81} have been investigated, in particular in the context of
the aging dynamics in spin glasses.
In this context also the function $\xi(t,t_w)$, which will be the central topic
of the present paper, has been discussed for the first time\cite{SH89, HS90}.
A well known model in this context is Bouchauds trap model\cite{trap} and a
number of investigations of the FDT violations have been
presented\cite{BD95,Ritort03,soll03}.

In the present paper, I consider models for which the probability
distributions obey a master equation (ME) and I assume that the transition
probabilities fulfill the conditions of detailed balance also in the presence
of a perturbing field. The behavior of the function $\xi(t,t_w)$ will be
discussed in detail for some specific models.
I will mainly focus on models, in which the dynamics of some stochastic
variable $M(t)$ is determined via transitions among the states of the system
under consideration.
Therefore, one has to assign values $M_k$ to the states $k$.
This can be done in a variety of different ways.
Here, I will discuss a class of models in which the stochastic process $M(t)$
is viewed as a projection from a two-dimensional composite Markov
process (CMP)\cite{vkamp81}.
When the transition rates are chosen in an appropriate way, this procedure
allows to treat several different models on the same footing,
including random magnetic models.
The outline of the paper is the following.
In the next section the general formalism will be discussed.
This includes a discussion of trap models because one generally finds that
$\xi(t,t_w)$ vanishes for these models\cite{BD95,Ritort03,soll03}.
Section III is devoted to random magnetic models and a particular example of
a kinetic random energy model is discussed in detail for illustrative
purposes. Models for tagged particle motion as they have been applied to the
dynamics in supercooled liquids are considered in Section IV and the
conclusions are presented in Section V.
\section*{II. Master equations and FDT violations}
\subsection*{A. General formalism}
Throughout this paper a stochastic dynamics according to a master equation
(ME)\cite{vkamp81} is assumed.
In a discrete notation let $G_{kl}(t)$ be the conditional probability to find
the system in 'state' $k$ at time $t$ provided it was in 'state' $l$ at time
$t\!=\!0$. At this point it is not necessary to specify the meaning of the term
'states'.
Denoting the transition probabilites for a transition from state
$k$ to state $l$ by $W_{lk}$, the ME reads as:
\be\label{ME}
\dot{G}_{kl}(t)=-\sum_nW_{nk}G_{kl}(t)+\sum_nW_{kn}G_{nl}(t)
\ee
Of course, the same ME is obeyed by the populations of state $k$,
$p_k(t)=\sum_lG_{kl}(t)p_l(0)$.
In addition, I only consider transition probabilities that obey detailed
balance
\be\label{det.balance}
W_{kl}p_l^{eq}=W_{lk}p_k^{eq}
\ee
where the $p_k^{eq}$ are the populations in thermal equilibrium.
These are invariant with respect to the master operator, i.e.
$p_k^{eq}=\sum_lG_{kl}(t)p_l^{eq}$.

Throughout the present paper it is assumed that the system is prepared in some
initial state described by a fixed set of populations, $p_k^0$.
The populations then evolve according to $p_k(t)=\sum_lG_{kl}(t)p_l^0$.
In the general case, of course, the $p_k^0$ are different from the $p_k^{eq}$,
but they still fulfill the sum rule $\sum_kp_k^0\!=\!1$ as required for
probabilities.

Central to the topic of the present paper is the two-time correlation function
\be\label{C.t.tw}
C(t,t_w)=\lg M(t)M(t_w)\rg=\sum_{k,l}M_kM_lG_{kl}(t-t_w)p_l(t_w)
\ee
where $M_k$ is the value of $M(t)$ in state '$k$'.
In this expression, $t_w$ denotes the time that has evolved after the initial
preparation of the system in the populations $p_k^0$.
In the following $t\!\geq\!t_w$ will always be assumed.
In addition, the correlation function can always be decomposed according to
\be\label{C.Ceq.DC}
C(t,t_w)=C_{eq}(t-t_w)+\D C(t,t_w)
\quad\mbox{with}\quad
C_{eq}(t-t_w)=\sum_{k,l}M_kM_lG_{kl}(t-t_w)p_l^{eq}
\ee
because of the properties of $G_{kl}(t_w)$. Note that this property only holds
if the $W_{kl}$ obey detailed balance, cf. eq.(\ref{det.balance}).
From $G_{kl}(t_w\!\to\!\infty)\!=\!p_k^{eq}$ one sees that $\D C(t,t_w)$
vanishes for long waiting times, because exactly this long time limit has been
subtracted from $\D C(t,t_w)$.
Eq.(\ref{C.Ceq.DC}) holds for all models considered in the present paper.

If $\lg M(t)\rg\!\neq\!0$, it is advantageous to consider
$\hat C(t,t_w)\!=\!\lg M(t)M(t_w)\rg -\lg M(t)\rg\lg M(t_w)\rg$, given by
$\hat C(t,t_w)\!=\!\sum_{k,l}M_kM_l\left[G_{kl}(t-t_w)-p_k(t)\right]p_l(t_w)$.

In order to calculate the linear response of the system,
\be\label{R.def}
R(t,t_w)=\left.{\d\lg M(t)\rg\over\d H(t_w)}\right|_{H=0}
\ee
the dependence of the transition probabilities on a field $H$ conjugate to $M$
has to be fixed.
In principle there is no restriction regarding this dependence.
From equilibrium considerations one expects a Boltzmann-like dependence,
$e^{\b HM}$. This, however, does not fix the dependence on the values of $M_k$
in the initial or final state of a $k\to l$ transition.
In the present paper, I choose the following form, which assures that
the system also in the presence of the field fulfills detailed balance:
\be\label{kap.h}
W_{kl}(H)=W_{kl}e^{\b H X_{kl}}
\quad\mbox{with}\quad X_{kl}=\a M_k-(1-\a)M_l
\ee
Here, $\a$ is a parameter that can take on any value.
The same dependence has also been used by Bouchaud and Dean\cite{BD95} in a
study of the aging properties of the trap model\cite{trap}.
Very recently, Ritort has generalized this dependence in using
$X_{kl}=\g M_k-\mu M_l$ with arbitrary $\g$ and $\mu$ with the effect that the
$W_{kl}(H)$ do no longer obey detailed balance\cite{Ritort03}.

Even though $\a$ in principle can take on any value, there often will be some
guiding principle. For example, if the states '$k$' denote the energies in a
canonical ensemble, one expects that $\a$ can be determined from the
dependence of the unperturbed $W_{kl}$ on $k$ and $l$. If the $W_{kl}$ are of
a form allowing a Kramers-Moyal expansion\cite{risken}, and therefore the ME
has a well defined Fokker-Planck equation as a limit one would naturally
choose $\a\!=\!1/2$.
However, it has to be pointed out that usually the states $k$ are understood
as metastable states or components\cite{palmer82} in connection with glassy
systems. Then the corresponding free energies are to be viewed as
coarse-grained quantities\cite{CR03} and one does no longer have a strict
relation of $\a$ to the unperturbed $W_{kl}$.

Using eq.(\ref{kap.h}), the linear response $R(t,t_w)$, eq.(\ref{R.def}), can
be calculated with the result
\be\label{R.t.tw}
R(t,t_w)=\b\sum_{k,l,n}M_k\left[G_{kn}(t-t_w)-G_{kl}(t-t_w)\right]
				 W_{nl}X_{nl}p_l(t_w)
\ee
After some algebra this can be related to the correlation function:
\be\label{FDT.gen}
R(t,t_w)=\b\left[ \a{\partial C(t,t_w)\over\partial t_w}
		 -(1-\a){\partial C(t,t_w)\over\partial t}
		 +\a\xi(t,t_w)\right]
\ee
where I defined the function
\be\label{xi.def}
\xi(t,t_w)=\sum_{k,l,n}M_kM_lG_{kn}(t-t_w)
		     \left[W_{nl}p_l(t_w)-W_{ln}p_n(t_w)\right]
\ee
It is important to point out that eq.(\ref{FDT.gen}) holds for arbitrary
Markov processes obeying detailed balance.
Additionally, the function $\xi(t,t_w)$ cannot be related to a time derivative
of the correlation function and therefore the response is not determined by
$C(t,t_w)$ alone in the general case. $\xi(t,t_w)$ plays a similar role as the
asymmetry in the treatment of the response derived from a Langevin
equation\cite{CKP94}.
Of course, a relation between $R(t,t_w)$ and $\hat C(t,t_w)$ mentioned above
is easily obtained from eq.(\ref{FDT.gen}) by noting that
$\partial\hat C(t,t_w)\!/\!\partial t\!=\!\partial C(t,t_w)\!/\!\partial t
-\partial(\lg M(t)\rg\lg M(t_w)\rg)\!/\!\partial t$ and similarly for
$\partial\hat C(t,t_w)\!/\!\partial t_w$.

If the system is prepared in an equilibrium state initially,
$p_k^0\!=\!p_k^{eq}$, one has $p_l(t_w)=p_l^{eq}$ and eq.(\ref{det.balance})
shows that $\xi_{eq}(t,t_w)\!\equiv\!0$. Furthermore, the response and the
correlation are functions of the time-difference only, i.e.
$C_{eq}(t,t_w)\!=\!C_{eq}(t-t_w)$ and $R_{eq}(t,t_w)\!=\!R_{eq}(t-t_w)$ and
thus
\be\label{fdt.eq}
R_{eq}(t)=-\b{d C_{eq}(t)\over d t}
\ee
which is just the well known FDT.

To the best of the authors knowledge, eq.(\ref{FDT.gen}) has not been
derived in this form before.
However, equations similar to eq.(\ref{FDT.gen}) have been given for various
models in the literature. Hoffmann and Sibani\cite{HS90} have derived
eq.(\ref{FDT.gen}) for the special case $\a\!=\!1$.
Later on Bouchaud and Dean\cite{BD95} give a similar expression with, however,
$\xi(t,t_w)\!=\!0$. Furthermore, Fielding and Sollich\cite{FS02} derived
eq.(\ref{FDT.gen}) for the special case of $\a\!=\!0$.
In all these cases the authors considered model systems where the quantities
$M_k$ denote a magnetization which is attached to state $k$ in some way.
Such models will be treated as special cases in the following chapters.
Before doing so it is instructive to have a somewhat closer look at the
quantity $\xi(t,t_w)$.
As noted by Hoffmann and Sibani, $\xi(t,t_w)$ vanishes if the relaxation to
equilibrium is determined by distribution functions that are equilibrated with
respect to the states $k$ and depend on time only parametrically.
In order to see this explicitly let us assume that the initial conditions are
such that one can write $p_k(t_w)=p_k^{eq}\d_k(t_w)$ with unspecified
functions $\d_k(t_w)$.
In this case eq.(\ref{xi.def}) reads as
\[
\xi(t,t_w)=\sum_{k,l,n}M_kM_lG_{kn}(t-t_w)W_{nl}p_l^{eq}
		     \left[\d_l(t_w)-\d_n(t_w)\right]
\]
From this expression it is evident immediately that $\xi(t,t_w)$ vanishes for
$\d_k(t_w)\!=\!\d(t_w)$ $\forall k$.

In the following sections some specific choices of the states $k$ and of the
dynamic variables will be presented and eq.(\ref{FDT.gen}) will be discussed
for these cases. In particular, I will concentrate on the function
$\xi(t,t_w)$ which according to the above discusssion is a measure of
inhomogeneities in the relaxation process under consideration.
\subsection*{B. Trap models}
As already mentioned above, Bouchaud and Dean\cite{BD95} derived
eq.(\ref{FDT.gen}) for a trap model, although with $\xi(t,t_w)\!=\!0$.
In order to understand this, one has to consider the quantities of interest in
the trap model in more detail.
Instead of the correlation function given in eq.(\ref{C.t.tw}), Bouchaud and
Dean consider a slightly different function:
\[\tilde C(t,t_w)=q_{EA}\Pi(t,t_w)\]
where $\Pi(t,t_w)$ denotes the propability that the system has not left the
initial trap during $t+t_w$ and $q_{EA}$ is the Edwards-Anderson order
parameter\cite{BY86}.
Note that $\Pi(t,t_w)$ can be interpreted as a structure factor for large
wave-vectors, because every escape out of a given trap gives rise to a
decorrelation and thus to a decay of $\Pi(t,t_w)$.
If one now identifies the unspecified states $k$ of the last section with the
traps and neglects any correlation between the values $M_k$ and the trap
energies, only the $G_{kk}(t)$ are relevant\cite{soll03}.
Assuming that the $M_k$ are distributed among the traps according to some fixed
distribution $\rho(M)$, one finds for the correlation function
\[
C(t,t_w)=\Pi(t,t_w)=\lg M^2\rg\sum_kG_{kk}(t-t_w)p_k(t_w)
\]
with $q_{EA}=\lg M^2\rg$ and accordingly
\[
\tilde\xi(t,t_w)=\lg M\rg^2\sum_{k,l}G_{kk}(t-t_w)
		     \left[W_{kl}p_l(t_w)-W_{lk}p_k(t_w)\right]
\]
From this last expression it is evident that for a symmetric distribution
$\rho(-M)\!=\!\rho(M)$ $\tilde\xi(t,t_w)$ vanishes, $\tilde\xi(t,t_w)\!=\!0$,
and the relation between the response and the correlation reads as
\[
\tilde R(t,t_w)=\b\left[ \a\partial_{t_w}\tilde C(t,t_w)
		 -(1-\a)\partial_t\tilde C(t,t_w)\right]
\]
For $\a\!=\!0$, i.e. a dependence only on the initial trap, this expression
reduces to the one given by Bertin and Bouchaud\cite{BB02}.
Very recently, Sollich\cite{soll03} has shown that the above expression holds
for any trap model independent of the form of the transition rates.
It will be shown later that $\xi(t,t_w)\!=\!0$ is also found for other models
in which any transition yields a complete decorrelation of the stochastic
variable of interest.
\subsection*{C. Composite Markov processes}
In this section I will consider the following class of models.
In order to relate the quantities $M_k$ to the values of $k$, I now treat the
states as configurations of the system, characterized by their (free)
energies $\e_k$ and consider the composite Markov process (CMP)
$\{\e(t),m(t)\}$\cite{vkamp81}.
Typically, one is interested in the dynamics of the process $m(t)$ only.
For example, when considering the relaxation properties in a (free) energy
landscape determined by some model, often a random assignment of the $m_k$ to
the states $k$ is used.
The stochastic process $m(t)$ is obtained from the CMP $\{\e(t),m(t)\}$
via the definition of marginal probabilities, i.e. by integrating out
$\e(t)$\cite{vkamp81}.
Note, that usually the process $m(t)$ defined this way is no longer a
Markov process.
Generally, $m(t)$ can be related to the quantities $M_k$ in various ways.
For instance, in the magnetic models to be treated in the following section
one identifies the $M_k$ with 'magnetizations' $m_k$.
Another example is given by the case of the reorientation of a molecule in a
liquid and the application of an electric field.
In this case one chooses $M_k\!=\!\mu\cos{(\Om_k)}$ with $\Om_k$ denoting the
orientation of the molecule and $\mu$ the value of the static dipole moment.
Assuming that $\Om$ is a random variable, one considers the CMP
$\{\e(t),\Om(t)\}$. Similarly, one can treat the translational motion of
tagged particles by identifying $M_k$ with $e^{i{\bf q}{\bf r}_k}$, where
${\bf r}_k$ denotes the position and now the CMP $\{\e(t),{\bf r}(t)\}$ is
considered.

The CMP $\{\e(t),m(t)\}$ is completely specified by the transition rates
$W(\e_k,m_k|\e_l,m_l)$. Of course, there are many ways to choose these rates
and in the present paper I will make the specific assumption:
\be\label{kap.gen}
W(\e_k,m_k|\e_l,m_l)=W(m_k|m_l) + \k_{kl}\L_{(kl)}(m_k|m_l)
\ee
Here, the $W(m_k|m_l)$ denote transition rates for a $m_l\!\to\!m_k$
transition and the $\k_{kl}$ those for a $\e_l\!\to\!\e_k$ transition.
Finally, the $\L_{(kl)}(m_k|m_l)$ determine what kind of change takes place
with the $m_k$ in case of a $\e_l\!\to\!\e_k$ transition.
The choice made in eq.(\ref{kap.gen}) is general enough to cover most
situations of interest.
Furthermore, it will be assumed throughout that no correlation between the
$\e_k$ and the $m_k$ exists initially and therefore the corresponding
probabilities factorize:
\be\label{p0.e.m.in}
p(\{\e_k,m_k\};t\!=\!0)=p(\e_k;0)p_k^{eq}(m_k)=p_k^0p_k^{eq}(m_k)
\ee
This means that additionally only the states $k$ are affected by the initial
non-equilibrium situation. This choice is sufficient for all models considered
in the present paper.

The two classes of models that will be considered in the present paper are
completely determined by the choice of the $W(m_k|m_l)$ and the
$\L_{(kl)}(m_k|m_l)$.
Random magnetic models can be obtained from eq.(\ref{kap.gen}) with the choice
\be\label{CMP.RM}
W(m_k|m_l)=0 \quad\mbox{and}\quad\L_{(kl)}(m_k|m_l)=\s_k(m_k)
\ee
where $\s(m_k)$ denotes the a priori distribution of the $m_k$ in
state $k$, usually assumed to be independent of $k$, $\s_k(m_k)\!=\!\s(m_k)$
$\forall k$.
Note that $\L_{(kl)}(m_k|m_l)\!=\!\s_k(m_k)$ means that this
quantity solely depends on the destination state.
The implications of eq.(\ref{CMP.RM}) will be worked out in the next section.

Another class of models concerns the models for dipole reorientation and
translational noted already above.
Here, one considers the CMP $\{\e(t),a(t)\}$, with $a(t)\!=\!\Om(t)$ or
$a(t)\!={\bf r}(t)$.
In this example, eq.(\ref{kap.gen}) reads as:
\be\label{CMP.Om}
W(\e_k,a|\e_l,a')=W_k(a|a')\d_{k,l} + \k_{kl}\L(a|a')
\ee
where I have already neglected a possible dependence of $\L$ on $k,l$.
Two classes of models can be distinguished in the present context.
One class is given by the so-called environmental fluctuation models
and is defined by the choice of the on-site relaxation rates $W_k(a|a')$
and of $\L(a|a')$. Usually, the latter quantity is either set to unity
(no particle movement associated with an exchange) or one chooses
$\L(a|a')\!=\!p^{eq}(a)$ (complete randomization).
In case of rotational motion the latter choice means that with any exchange
process a random reorientation takes place\cite{BP65,sill96,AU67}.

The 'energy landscape model' is defined by assuming that there is no
on-site relaxation at all, $W_k(a|a')\!=\!0$, and the geometry of the
molecular motion is determined by $\L(a|a')$\cite{dieze97}.
Note that in this model the process $\e(t)$ really defines the dynamics and
$a(t)$ can be viewed as some kind of slave process.
In its applications to the dynamics of supercooled liquids the assumption of
angular jumps of a finite width $\D\Om$\cite{DS99}, i.e.
$\L(\Om|\Om')\!=\!\d(\Om-[\Om'+\D\Om])$, has proven successful\cite{DSHB98},
but also random rotational jumps, $\L(\Om|\Om')\!=\!p^{eq}(\Om)$, have been
used for general considerations\cite{dieze97}.
In case of translational motion of tagged particles, an isotropic model with
$\L({\bf r}|{\bf r}')\!\propto\!\d(r-[r'+\d R])$ with a jump length $\d R$ has been
used to give a simple explanation of the apparent translational enhancement in
supercooled liquids\cite{DSHB98}.
Before considering these kinds of models further, I will treat the random
magnetic models in the following section.
\section*{III. Random magnetic models}
In the usual treatment of magnetic models one simply assumes that a random
magnetization $m_k$ is assigned to the states '$k$'\cite{HS90,FS02}.
Here, I will treat the CMP $\{\e(t),m(t)\}$ as introduced in the
preceeding section with the transition rates determined by eq.(\ref{CMP.RM}),
i.e.
\be\label{kap.rm}
W(\e_k,m_k|\e_l,m_l)=\s_k(m_k)\k_{kl}
\ee
The initial condition, eq.(\ref{p0.e.m.in}), for these models reads as
$p(\e_k,m_k;0)\!=\!p_k^0\s_k(m_k)$.
Starting from the ME, eq.(\ref{ME}), where now $\sum_k$ has to be read as
$\sum_k\int\!dm_k$, one finds that the ansatz
$G(\{\e_k,m_k\},t|\{\e_l,m_l\},0)\!=\!\s_k(m_k)G_{kl}(t)$ with
$G_{kl}(t)\!\equiv\!G(\e_k,t|\e_l,0)$ solves the ME. The same applies for the
time dependent probabilities, $P(\{\e_k,m_k\},t)\!=\!\s_k(m_k)P_k(t)$.
In calculating the correlation, the response and $\xi(t,t_w)$ some care has to
be taken in performing multiple summations. For instance, in an
abbreviated form one has
$\xi(t,t_w)\!=\sum_{k,l,n}\int\!dm_k\int\!dm_l\int\!dm_l
\s_k(m_k)\s_l(m_l)\s_n(m_n)m_km_lA_{k,l,n}$ after all substitutions have been
performed.
In such an expression, one has to treat the terms $k\!=\!l$ and $k\!\neq\!l$
separately and then perform the integrations.
This way all quantities of interest can be obtained from
eqns.(\ref{C.t.tw},\ref{R.t.tw},\ref{FDT.gen},\ref{xi.def}).
In the most general case the results depend on the moments
$\lg m_k^n\rg\!=\!\int\!dm_k\s_k(m_k)m_k^n$. For example, the correlation is
found to be given by:
\be\label{C.rm.gen}
C(t,t_w)=\sum_k\lg\D m_k^2\rg G_{kk}(t-t_w)p_k(t_w)
	 +\sum_{k,l}\lg m_k\rg\lg m_l\rg G_{kl}(t-t_w)p_l(t_w)
\ee
with $\lg\D m_k^2\rg\!=\!\lg(m_k-\lg m_k\rg)^2\rg$.
If even distribution functions, $\s_k(m_k)\!=\!\s_k(-m_k)$, for which
$\lg m_k\rg\!=\!0$ are chosen, the second term in the above expression for the
correlation function vanishes.
For $\xi(t,t_w)$, one finds in this case:
\be\label{xi.rm.m0}
\xi(t,t_w)=\sum_{k,l}\lg m_k^2\rg G_{kl}(t-t_w)
		     \left[\k_{lk}p_k(t_w)-\k_{kl}p_l(t_w)\right]
\ee
These expressions can be analyzed for arbitrary random magnetic models.
It has, however, to be pointed out that the simple choice, eq.(\ref{kap.rm})
is meaningful only, if the magnetizations are randomly distributed among the
energies, because this cannot account for any correlations between
the $m_k$ and the $\e_k$.
In the following, I will illustrate the general formulae for the case of a
simple model for which the master equation can be solved analytically.
\subsection*{Example: kinetic random energy model}
As an example, I will treat the special case of a kinetic random energy
model (REM) with a special choice of the transition rates $\k_{kl}$.
I will follow the treatment of Koper and Hilhorst\cite{KH89},
who considered various forms of the transition rates.
For the purpose of the present paper the simplest choice
\be\label{k.kl.rcm}
\k_{kl}=\k_0\exp{(-\b\e_k)}
\ee
is sufficient, where $\k_0$ is a rate constant, to be set to unity in the
following, and $\b\!=\!T^{-1}$.
This means that the $\k_{kl}$ only depend on the destination state of the
transition via the Boltzmann factors $B_k=\exp{(-\b\e_k)}$.
The corresponding ME can easily be solved giving
$G_{ik}(t)\!=\!Z(\b)^{-1}B_i+\left[\d_{ik}-Z(\b)^{-1}B_i\right]\exp{(-Z(\b)t)}$
with the partition function $Z(\b)\!=\!\sum_k B_k$ and $\d_{ik}$ denotes
the Kronecker symbol. The equilibrium populations are given by
$p_i^{eq}\!=\!Z(\b)^{-1}B_i$.

In principle one can choose the random variable to depend on the $\e_k$, a
possible choice being $\lg m_k^2\rg\!=\!B_k^n$ for arbitrary $n$, cf. the
discussion given on this point by Fielding and Sollich\cite{FS02}. Without
showing explicit results here for the case $n\!\neq\!0$, it is appropriate
to mention that also for the simple model considered in this section, the
FDT-violations depend on the value of $n$, meaning that the $m_k$ chosen this
way do not represent neutral variables\cite{CR03,FS02}. In the following, I will
choose $n\!=\!0$, $\lg m_k\rg\!=\!0$  and $\lg m_k^2\rg\!=\!1$.
In addition, the initial populations $p_i^0$ will be chosen as
$p_i^0\!=\!N^{-1}$, appropriate for a quench from high temperatures in the
beginning of the experimental protocol.
For this case one finds, using the abbreviation
${\cal Z}(\b)\!=\!Z(2\b)/Z(\b)^2$:
\Be\label{C.rcm.ungem}
C^{(B)}(t,t_w)=&&\hspace{-0.6cm}
C_{eq}^{(B)}(t-t_w)+\D C^{(B)}(t,t_w)\nonumber\\
C_{eq}^{(B)}(\t)=&&\hspace{-0.6cm}
{\cal Z}(\b) +\left[1-{\cal Z}(\b)\right]\exp{(-Z(\b)\t)}\\
\D C^{(B)}(t,t_w)=&&\hspace{-0.6cm}
{\cal Z}(\b)\left[\exp{(-Z(\b)t)}-\exp{(-Z(\b)t_w)}\right]\nonumber
\Ee
and
\be\label{Ksi.rcm.ungem}
\xi^{(B)}(t,t_w)=Z(\b)\D C^{(B)}(t,t_w)
\ee
For the response one has
\be\label{R.rcm.ungem}
R^{(B)}(t,t_w)=-\b\left[{\partial C_{eq}^{(B)}(t-t_w)\over\partial t}
		  -Z(\b){\cal Z}(\b)\exp{(-Z(\b)t)}\right]
\ee
In these expressions, the explicit dependence on the Boltzmann factors $B_i$
has been denoted by the superscript $B$.
Interestingly, the expression for the response is independent of $\a$.
It should, however, be noted that in the more general case of a variable
$\lg m_k^2\rg\!=\!B_k^n$, $n\!\neq\!0$, $R^{(B)}(t,t_w)$ is found to depend
on $\a$.

Below the transition temperature $T_c\!=\!J/(2\sqrt{\log{2}})$ the random
energies are exponentially distributed\cite{KH89}, where $J$ is related to
the variance of the Gaussian distribution of the REM\cite{derrida80}.
The corresponding distribution of Boltzmann factors $B_k$ is given by
\be\label{p.Bk}
p(B)={\nu\over N}x B^{-1-x}\quad\mbox{for}\quad
      \left({\nu\over N}\right)^{(1/x)}<B<\infty
\ee
and $p(B)\!=\!0$ otherwise. Here, $x\!=\! T/T_c$,
$\nu\!=\!1/(2\sqrt{\pi\log{2}})$ and $N$ denotes the number of random
energies.
All quantities of interest are obtained as averages over the distribution
$p(B)$, i.e.
\be\label{F.av}
F(t)=\lg F^{(B)}(t)\rg_B=\int\!dBp(B)F^{(B)}(t)
\ee
In order to perform the averages of the quantities given in
eqns.(\ref{C.rcm.ungem},\ref{Ksi.rcm.ungem},\ref{R.rcm.ungem}), the only
integrals required are:
\Be\label{Phi.Psi.def}
&&\Phi(t)=\lg e^{-Z(\b)t}\rg=\exp{(-\tilde{v}t^x)}\nonumber\\
&&\Psi(t)=\lg{\cal Z}(\b)e^{-Z(\b)t}\rg
=(1-x)\left[\Phi(t)-\tilde{v}^{1/x}\G(1-1/x;\tilde{v}t^x)\right]
\Ee
where $\tilde{v}\!=\!v\G(1-x)$ with the Gamma function $\G(a)$. Furthermore,
$\G(a,b)$ denotes the incomplete Gamma function. The calculation of the
averages follows the lines of ref.\cite{KH89} and poses no problem.

In particluar, one finds:
\Be\label{C.R.rcm}
C_{eq}(\t)=&&\hspace{-0.6cm}
1-x+\Phi(\t)-\Psi(\t)\nonumber\\
\D C(t,t_w)=&&\hspace{-0.6cm}
\Psi(t)-\Psi(t_w)\\
R(t,t_w)=&&\hspace{-0.6cm}
-\b\partial_t\left[C_{eq}(t-t_w)+\Psi(t)\right]\nonumber
\Ee
and
\be\label{Ksi.rcm}
\xi(t,t_w)=\partial_{t_w}\Psi(t_w)-\partial_t\Psi(t)
\ee
Koper and Hilhorst\cite{KH89} treated $C_{eq}(\t)$ in their investigation.
Using the asymptotic behavior of $\G(a,b)$ it can be shown
that this function behaves as
$C_{eq}(\t)\!\sim\!\exp{(-\tilde{v}t^x)}$ at long times.
For the integrated response $\chi(t,t_w)\!=\!\int_{t_w}^t\!dt'R(t,t')$
one finds:
\be\label{Chi.t.tw}
\chi(t,t_w)=\b\left[C_{eq}(0)-C_{eq}(t-t_w)-(t-t_w)\partial_t \Psi(t)\right]
\ee
where the first two terms correspond to the equilibrium response,
$\chi_{eq}(\t)=\b\left[1-C_{eq}(\t)\right]$, because $C_{eq}(0)\!=\!1$ for
$n\!=\!0$.
The limiting behavior of the integrated response can be summarized as follows:
\Be\label{chi.limit}
t_w\to\infty:\quad\chi(t_w+\t,t_w)\to&&\hspace{-0.6cm}
\chi_{eq}(\t)\quad\hspace{1.3cm}\forall\t\nonumber\\
\t\to\infty:\quad\chi(t_w+\t,t_w)\to&&\hspace{-0.6cm}
\b x\equiv T_c^{-1}\quad\hspace{0.65cm}\forall t_w\\
t_w\to 0:\quad\chi(t_w+\t,t_w)\to&&\hspace{-0.6cm}
\b x\left[1-\Phi(\t)\right]\quad\forall\t\nonumber\\
\t\to 0:\quad\chi(t_w+\t,t_w)\to&&\hspace{-0.6cm}
0\quad\hspace{2.2cm}\forall t_w\nonumber
\Ee
This behavior is illustrated in Fig.1, which shows a typical
FDT-plot\cite{CKP97}, $\b^{-1}\chi(t_w+\t,t_w)$ versus $C(t_w+\t,t_w)$,
for $x\!=\!0.3$.
From this plot and eq.(\ref{chi.limit}) it is evident that the limiting
fluctuation-dissipation ratio\cite{CR03} for this model is
given by $X_\infty\!=\!T_c^{-1}$. The dotted lines in Fig.1 represent
a slope of ($-1$) (ordinary FDT) and a slope of ($-x$), the limiting behavior
for long times $\t$.
Therefore, in this simple model, the effective temperature coincides with the
transition temperature $T_c$.

The function $[-\xi(t_w+\t,t_w)]$ is plotted versus $\t$ in Fig.2 for various
values of the waiting time $t_w$ for $x\!=\!0.3$. The behavior for other
values of $x$ is very similar.
For very short $\t$ the function $\xi$ starts from zero and approaches a
plateau value determined by $[\partial_{t_w}\Psi(t_w)]$ for long times $\t$,
cf. eq.(\ref{Ksi.rcm}). Only in the limit of long waiting one has
$\xi(t_w+\t,t_w)\!=\!0$.
The lower panel of Fig.2 shows a logarithmic plot of $(-\xi)$, which
demonstrates the linear behavior of $\xi$ and also that the cross-over to the
plateau value happens around the waiting time, i.e. for $\t\sim t_w$.
For not too long waiting times, the approximate expression
\be\label{Ksi.rcm.ap}
\xi(t_w+\t,t_w)\simeq (x\tilde{v})t_w^{x-1}
	    \left[\left(1+{\t\over t_w}\right)^{x-1}-1\right]
\ee
holds with high accuracy. This expression is shown in the lower panel of Fig.2
as the thin dotted lines. For $t_w\!=\!10^{-5}$, it cannot be distinguished
from the exact expression, eq.(\ref{Ksi.rcm}).

From these considerations it becomes clear that even for this simple example
the assumption that the distribution function determining the relaxation
is equilibrated with respect to the $\e_k$ is meaningless.
On the contrary, the system gets stuck in a distribution characterized solely
by the transition temperature $T_c$.
In order to further discuss the properties of the function $\xi(t,t_w)$, in
the next section I consider a different class of models.
\section*{IV. Models for tagged particle motion}
Models exhibiting truely non-equilbrium dynamics are often considered as models
for glassy systems. However, when one is concerned with supercooled liquids,
the usual situation is one in which the system can reach metastable
equilibrium for long times. The glass transition temperature is merely a
convenient way of classifying the time scale of the primary
relaxation\cite{EAN96}.
Nevertheless, investigating the out-of-equilibrium dynamics of simple models
for molecular motion in supercooled liquids is capable of providing additional
information about the mechanisms of the dynamics.

In this section, I consider models for tagged particle motion that have
been used successfully to describe the translational and reorientational
dynamics in supercooled liquids on a phenomenological
level\cite{sill96,dieze97,DSHB98}.

In case of the reorientational motion, the correlation function $C(t,t_w)$,
eq.(\ref{C.t.tw}), is obtained by identifying $M(t)$ with orientation
dependent static dipole moments $\mu(t)$.
The time dependence has its origin in the temporal fluctuations of the
direction of $\mu$ relative to a space fixed axis.
It should be pointed out that $C(t,t_w)$ is directly related to the so
called rotational correlation function $g_1(t,t_w)$, where the general
definition is
\be\label{gl.t}
g_L(t,t_w)=\lg P_L(\cos{(\theta(t))}P_L(\cos{(\theta(t_w))}\rg
\ee
Here, the angle $\theta$ is the polar angle of the axially symmetric principal
axes system in the laboratory fixed coordinate system and $P_L(x)$ denotes
the Legendre polynomial of rank $L$.
For instance, in dielectric spectroscopy $\theta$ is the angle between the
applied field and $\mu$ and $g_1(t,t_w)$ is measured. In NMR or light
scattering $g_2(t,t_w)$ typically is observed.

If one is interested in translational motions of tagged particles, one choice
is to identify $M(t)$ with $e^{i{\bf q}{\bf r}(t)}$, with ${\bf q}$ denoting
the wave-vector and ${\bf r}(t)$ is the position of the particle.
This means, one has a close relation between $C(t,t_w)$ and the
incoherent intermediate scattering function\cite{squires}:
\be\label{Sq.t}
C^{(q)}(t,t_w)=\lg e^{i{\bf q}{\bf r}(t)}e^{-i{\bf q}{\bf r}(t_w)}\rg
\ee
where I have already assumed that the system is isotropic, $q\!=\!|{\bf q}|$.

In the following I consider the CMP $\{\e(t),a(t)\}$, where the $a(t)$
either denotes the orientation $\Om(t)$ or the position ${\bf r}(t)$.
\subsubsection*{Energy landscape models}
I start the discussion with the CMP $\{\e(t),a(t)\}$ in the context of the
energy landscape model introduced in ref.\cite{dieze97}.
In this model, one associates a (free) energy $\e_k$ with each basin or
metabasin of the potential energy landscape of the system\cite{CR03,SW83}.
In the model of a random first order transition put forward by Wolynes et
al., see e.g. ref.\cite{XW01}, the $\e_k$ would be interpreted as the free
energies of the metastable glassy states, characterized by an analogue of the
Edwards-Anderson order parameter.
The slow primary relaxation in such scenarios is determined by the transition
among the various states. In the energy landscape model a tagged particle is
allowed to reorient about an average jump angle $\Theta$ and also to hop a
mean distance $\d\!R$ whenever a $k\!\to\!l$ transition takes
place\cite{dieze97,DSHB98}.
As already mentioned in Sect.IIC, the simplest choice for the transition rates
given in eq.(\ref{CMP.Om}) is $W_k(a|a')\!=\!0$ and
$\L(a|a')\!\propto\!\d(a-[a'+\D a])$ with $\D a$ denoting $\Theta$ or $\d\!R$.
In the context of reorientational motions, sometimes it is sufficient to
consider rotational random jumps, which just means that any $k\!\to\!l$
transition randomizes the molecular orientations completely.
In this case one chooses $\L(\Om|\Om')\!=\!1/(8\pi^2)$.

In Appendix A the solution of the ME for this model is explained and the
expressions for the quantities relevant in the present context are derived.
The correlation function $C(t,t_w)$ is given in eq.(\ref{C.p}) and $\xi(t,t_w)$
in eq.(\ref{xi.p}).
From the latter expression it is obvious that $\xi$ is closely related to the
non-equilibrium part of the correlation function, $\D C$.
The response then follows from the general expression, eq.(\ref{FDT.gen}).
For $\xi(t,t_w)$ it is found that it is given by
\[
\xi^{(1)}(t,t_w)\propto\cos{(\Theta)}
\]
in case of reorientations with an jump angle $\Theta$.
If random reorientations are assumed instead, one has
$\xi^{(1)}(t,t_w)\!\equiv\!0$.
For translational motion in an isotropic model one finds
\[
\xi^{(Q)}(t,t_w)\propto j_0(Q)
\]
where $j_0(x)$ is a Bessel function and $Q\!=\!(q\cdot\d\!R)$ with the
wave-vector $q$.
The limiting behavior is found to be given by ($p\!\equiv\!Q$ or $L$):
\Be\label{xi.ELM.limit}
t_w\to\infty:\quad\xi^{(p)}(t_w+\t,t_w)\to&&\hspace{-0.6cm}
0\quad\hspace{1.3cm}\forall\t\nonumber\\
\t\to\infty:\quad\xi^{(p)}(t_w+\t,t_w)\to&&\hspace{-0.6cm}
0\quad\hspace{1.3cm}\forall t_w\\
t_w\to 0:\quad\xi^{(p)}(t_w+\t,t_w)\to&&\hspace{-0.6cm}
\xi^{(p)}(\t,0)\quad\forall\t\nonumber\\
\t\to 0:\quad\xi^{(p)}(t_w+\t,t_w)\to&&\hspace{-0.6cm}
0\quad\hspace{1.3cm}\forall t_w\nonumber
\Ee
It is evident that in contrast to the simple kinetic REM considered in the
previous section, $\xi(t_w+\t,t_w)$ vanishes in the limit of long $\t$.
The reason for this difference is that in the present model the system
reaches equilibrium in this limit.

Although the general expressions for $\xi$ are not very instructive,
two limits can be analyzed in simple terms and the relevant formulae are
given in Appendix A.
One limit is given by either random rotations or large wave-vectors.
In both cases, one finds that $\xi$ vanishes. This means that whenever a
transition gives rise to a randomization of the relevant variable, $\xi$
is expected to vanish. This is very similar to the case ot the trap models
discussed in Sect.IIB.
Therefore, in this situation, the correlation function $C(t,t_w)$ is
essentially the same as the function $\Pi(t,t_w)$, giving the probability
for staying in the same orientation/position (as opposed to the same trap).
This can also be seen from the direct comparison of the expression for
$C^{(1)}(t,t_w)$ and $C^{(Q)}(t,t_w)$ given in eq.(\ref{All.p0}) with the one
given for $\Pi(t,t_w)$ in Sect.IIB, because only $G_{kk}(t)\!=\!e^{-\k_kt}$
enters these expressions, cf. eq.(\ref{Gp.p0}).
The correlation functions in this limit are independent of $L$ or $Q$ because
every transition gives rise to a complete decorrelation.
Thus, there is a close link between the simple model considered here and trap
models concerning the vanishing of the function $\xi(t,t_w)$.
It should, however, be noticed that this correspondence relies only on the
'geometric' aspects and has nothing to do with the aging dynamics in the
models considered.
The important point regarding $\xi(t,t_w)$ is the fact that due to the
randomization of the relevant stochastic variable in each transition the
distribution of orientations/positions is homogeneous.
A very important difference between the model for translational model and
the trap model, however, is that the wave-vector is a quantity that can be
adjusted experimentally. Therefore, the limit of vanishing $\xi^{(Q)}$ can be
reached continuously in principle.

The other linit that can be investigated in detail is that of very small
rotational jump angles $\Theta$ (rotational diffusion) or small wave-vectors
$q$. In this case, one finds that the equilibrium correlation functions decay
approximately exponentially in time with diffusion coefficients related to the
averaged transition rate $\lg\k\rg\!=\!\sum_{k,l}\k_{kl}p^{eq}_k$, cf.
eq.(\ref{Ceq.p.eta})\cite{DSHB98}. In this limit, the prefactor of $\xi$,
$\cos{(\Theta)}$ or $j_0(Q)$, are approximately equal to unity.

The general expressions for the correlation function and the function $\xi$
are illustrated in Appendix A by a simple two-state model.
As is to be expected, $\D C$ and also $\xi$ vanish for equal transition rates,
$\k_{12}\!=\!\k_{21}$, implying $p^{eq}_1\!=\!p^{eq}_2$.
Also the transient nature of $\D C$ and $\xi$ is obvious from eq.(\ref{TS.p0})
and eq.(\ref{TS.p.eta}).
\subsubsection*{Environmental fluctuation models for molecular reorientations}
In order to show that the results obtained in the framework of the energy
landscape model qualitatively are the same also for other models, I will now
discuss the exchange models mentioned earlier.
For simplicity it is asumed that there are two 'states', one in which molecular
reorientation is fast, denoted by '$f$' in the following, and one in which
molecules reorient slowly, denoted by '$s$'\cite{sill96}.
The generalization to an arbitrary number of states poses no problem and has
been used to describe several features of the reorientational relaxation in
supercooled liquids, including dielectric relaxation\cite{AU67} and four
dimensional NMR experiments\cite{sill96,RHDBS96}.
If the exhange rates, $\k_{sf}$ and $\k_{fs}$, vanish, one has a static
heterogeneous dynamics in the sense that slow molecules remain slow forever.
The solution of the corresponding ME and the relevant results are given in
Appendix B\cite{sill96}, where it is shown that there are only slight
differences to the case of the energy landscape model.

For the case of no changes in molecular orientation due to $k\to l$
transitions, orientational relaxation takes place solely due to on-site
reorientations.
From the expression for $\D C(t,t_w)$, eq.(\ref{C.BP}), it becomes evident that
this function vanishes, if the reorientation rates in the two environments,
$\G_f$ and $\G_s$, are equal, i.e. $\D C(t,t_w)\!=\!\xi(t,t_w)\!=\!0$ for
$\G_s\!=\!\G_f$. The same holds if one has $\k_{sf}\!=\!\k_{fs}$.
Furthermore, it is evident that in this model the only relaxation during the
waiting time is given by exchange. For vanishing exchange rates
$\xi$ vanishes, but $\D C$ remains at its initial value for all waiting times.
The latter fact is easily understood because in this case there is no way for
equilibration of the $p_{s/f}^0$.
As in the energy landscape model, $\xi(t_w+\t,t_w)$ vanishes in the limits of
small and large $\t$ and has a maximum at intermediate $\t$ determined by the
effective decay rates.

Sofar, I have considered the two-state model in which an exchange process has
no effect on the molecular orientation.
However, the other extreme scenario, in which the molecular orientation
randomizes completely in case of a $k\!\to\!l$ transition, can easily be
treated in the same way.
It turns out that the only relevant difference to the model considered
above is that now $\xi(t,t_w)\!=\!0$, cf. eq.(\ref{xi.BP.RJ}).
In this case the orientational distribution is homogeneous at each instant of
time and the argument for a vanishing $\xi$ given in Sect. IIA applies.
\section*{V. Conclusions}
I have considered the fluctuation-dissipation relation for a general class of
stochastic models obeying a master equation.
I have chosen transition rates which in the presence of a field are disturbed
in a multiplicative way. Only cases for which detailed balance is obeyed with
and without field were considered.

The main result of the present paper is given in eq.(\ref{FDT.gen}) and shows
that in the general case the response cannot be related to time-derivatives of
the correlation function alone. Instead, a function $\xi(t,t_w)$
occurs additionally in the expression for the response, which cannot be
related to any physical quantity in a simple manner.
An exception to this finding is provided by trap models, where it has been
shown the $\xi(t,t_w)$ vanishes for even distributions.
Hoffmann and Sibani\cite{SH89,HS90} were the first who considered the function
$\xi(t,t_w)$ for the special case of $\a\!=\!1$ in the general expression for
the fluctuation dissipation relation, eq.(\ref{FDT.gen}).
However, they did not further discuss the meaning of this function apart from
the fact that it vanishes if the relaxation is determined by probability
distributions that are homogeneous with respect to the states of the
stochastic process under consideration.

In the present paper I have shown that this assumption is likely to be wrong
even for very simple minded models. As one example I considered a kinetic
random energy model with an extremely simple choice for the transition rates
in Sect.III. The limiting effective temperature was found to coincide with
the transition temperature of the REM, $T_{\rm eff}\!=\!T_c$.
It was found that even for this model $\xi(t_w+\t,t_w)\!\neq\!0$
for $t_w\!<\!\infty$. Due to the fact that the system never reaches equilibrium
in this model, one also has a non-vanishing long time limit
$\xi(\infty,t_w)$.

Furthermore, I have shown that for some models in which any transition leads
to a complete loss of correlation of the stochastic variable under study,
one finds that the function $\xi(t,t_w)$ vanishes.
This holds for trap models and also for models of molecular reorientations,
provided these reorientations proceed via random rotational jumps.
If the the energy landscape model is applied to translational motions of
tagged particles, one has
$\xi^{(Q)}(t,t_w)\!\propto\!j_0(Q)$, $Q\!=\!q\!\cdot\!\d\!R$, if the
intermediate scattering function $C^{(Q)}(t,t_w)$ is measured.
If $Q\!\gg\!1$, one therefore also finds that $\xi^{(Q)}$ vanishes.
In this example, the fact that $\xi$ vanishes thus relies on an experimentally
adjustable parameter and does not reflect a property of the transition rates.
It is well known, that a stimulated NMR echo experiment is the 'rotational
analogue' of the intermediate scattering function\cite{BDHR01}.
If this experiment is performed in a way that is analogous to
$C^{(Q\!\gg\!1)}(t,t_w)$, every single rotational jump leads to a decorrelation
and the random rotational jump model applies.
Also in case of the trap model any transition out of a trap leads to a loss of
correlation.
Thus, also $\Pi(t,t_w)$ corresponds to $C^{(Q\!\gg\!1)}(t,t_w)$.
Therefore, it is tempting to speculate that whenever an analogue of
$C^{(Q)}(t,t_w)$ in the large $Q$-limit is considered for $C(t,t_w)$ in the
sense that every transition gives rise to a complete decorrelation, the
function $\xi^{(Q)}(t,t_w)$ is expected to vanish.
This also means that whenever every transition among the states of a system
obeying a ME randomizes the stochastic variable under consideration,
I expect $\xi(t,t_w)\!=\!0$.

For all models considered in the present paper, it turns out that $\xi(t,t_w)$
is related to the non-equilibrium part of the correlation function,
$\D C(t,t_w)$.
Furthermore, the main difference between the kinetic REM and the models
considered for tagged particle motion consists in the fact that in the latter
models the system reaches equilibrium for long times.
It is found that $\xi(t_w+\t,t_w)$ (if it is non-vanishing) for these models
is of a transient nature and vanishes for short and long $\t$, in contrast to
the case of the kinetic REM, where
$\lim_{\t\to\infty}\xi(t_w+\t,t_w)\!\neq\!0$.

At the moment it is not clear, how the function $\xi(t,t_w)$ behaves for other
types of models. A study of kinetic Ising chains is under way.
Also, it would be interesting to investigate the behavior of $\xi(t,t_w)$ for
varying initial conditions.
In the present paper, eq.(\ref{p0.e.m.in}) was always assumed to hold.

To conclude, I have shown that for some general class of models the fluctuation
dissipation relation is determined by time-derivatives of the correlation
function and an additional function $\xi(t,t_w)$.
As noted earlier\cite{HS90}, $\xi(t,t_w)$ is a measure for deviations from
homogeneity in the distribution functions determining the relaxation.
This function is expected to vanish in situations where transitions among the
states considered lead to a randomization of the relevant stochastic variable,
meaning that in such a case the relevant distributions are homogeneous.
\subsection*{Acknowledgements:}
I thank U. H\"aberle and R. Schilling for fruitful discussions
on the topic. Part of this work was supported by the DFG under
Contract Di693/1-1.
\begin{appendix}
\section*{Appendix A: Energy landscape model}
\setcounter{equation}{0}
\renewcommand{\theequation}{A.\arabic{equation}}
In this Appendix I will give the relevant formulae that are needed for the
solution of the ME for the CMP $\{\e(t),a(t)\}$, where $a(t)$ has to be
identified with the orientation $\Om$ or the position ${\bf r}$, depending
on the situation.
Because the solution of the ME proceeds in the same way for both cases, the
treatment can be formulated quite generally.
The transition rates are given by
\be\label{W.eps.a}
W(\e_k,a|\e_l,a')=\k_{kl}\L(a|a')
\ee
which follows from eq.(\ref{CMP.Om}) for $W_k(a|a')\!=\!0$. Again, I neglected
possible dependencies of $\L(a|a')\!\propto\!\d(a-(a'+\D a))$ on the initial
and final states of the transition.
To solve the corresponding ME, eq.(\ref{ME}), for this case, one defines the
matrix of the eigenvectors of $\L(a|a')$, $U(a,p)$ corresponding to the
eigenvalue $\L(p)$, i.e.
$\L(p)\!=\!\int\!da\int\!da'U^{-1}(a,p)\L(a|a')U(a,p)$.
Next, the conditional probability is expanded in terms of these eigenvectors,
\be\label{G.eps.a}
G(\{\e_k,a\},t|\{\e_l,a_0\},0)=
\int\!dpU(a,p)G^{(p)}_{kl}(t)U^{-1}(a_0,p)
\ee
The Greens functions $G^{(p)}_{kl}(t)$ are then found from the solution of
\be\label{Gp.ME}
{\dot G^{(p)}}_{kl}(t)= - \k_{k}G^{(p)}_{kl}(t)
			  +\L(p)\sum_n\k_{kn}G^{(p)}_{nl}(t)
\ee
where the sum rule $\k_k\!:=\!\sum_{l\neq k}\k_{lk}$ was used for the diagonal
element.
The initial populations are chosen according to eq.(\ref{p0.e.m.in}), which
in the present context reads as:
\be\label{p0.eps.a}
p(\{\e_k,a\};0)=p^{eq}(a)p_k^0
\ee
If one considers isotropic rotational motions with a fixed mean jump angle
$\Theta$, one has to identify $a$ with $\Om$. In this case one has
$p^{eq}(a)\!=\!{1\over 8\pi^2}$,
$U(a,p)\!=\!\sqrt{{2L+1\over 8\pi^2}}D^{(L)}_{mn}(\Om)$,
$\L(p)\!=\!P_L(\cos{(\Theta)})$ and the integration over $p$ is now a sum
over $L$.

In case of translational jumps onto all postions of a sphere with a radius
$\d\!R$ (jump length), the corresponding substitutions in the general formulae
are $p^{eq}(a)=V^{-1}$ with $V$ denoting the volume.
Furthermore, one has $U(a,p)\!=\!{1\over\sqrt{V}}e^{i{\bf q}{\bf r}}$ and
$\L(p)=j_0(q\d\!R)$, where $j_0(x)$ denotes the Bessel function of zeroth
order.

In eq.(\ref{Gp.ME}) one has to identify $p\!=\!L$ for rotations and
$p\!=\!Q$ with $Q\!=\!(q\!\cdot\!\d\!R)$ for translations and thus
eq.(\ref{Gp.ME}) reads as:
\Be\label{Gl.Gq.ME}
&&\rm{Rotation:}\hspace{1.5cm}
\dot{G}^{(L)}_{kl}(t)=-\k_kG^{(L)}_{kl}(t)
		       +P_L(\cos{(\Theta)})\sum_n\k_{kn}G^{(L)}_{nl}(t)
		       \nonumber\\
&&\rm{Translation:}\hspace{1cm}
\dot{G}^{(Q)}_{kl}(t)=-\k_kG^{(Q)}_{kl}(t)
		       +j_0(Q)\sum_n\k_{kn}G^{(Q)}_{nl}(t)
\Ee
Note that due to $P_0(x)\!=j_0(0)\!=\!1$, there is one eigenvalue, $p\!=\!0$,
$\L(0)\!=\!1$.
Therefore, for $p\!=\!0$, from eqns.(\ref{Gp.ME}, \ref{Gl.Gq.ME}) the original
ME for the transitions in the energy landscape is obtained, i.e.
$G^{(0)}_{kl}(t)\!\equiv\!G_{kl}(t)$.
This is in full accord with the idea underlying the model that the variable $a$
is allowed to change {\it only} in case of a $k\!\to\!l$ transition among
the states of the system\cite{DSHB98}.

Once the $G^{(p)}_{kl}(t)$ are obtained from the (numerical) solution of the
equations (\ref{Gl.Gq.ME}), all quantities of interest can be calculated and
one finds, cf. eq.(\ref{C.Ceq.DC}):
\be\label{C.p}
C^{(p)}_{eq}(\t)=\sum_{k,l}G^{(p)}_{kl}(\t)p^{eq}_l
\quad;\quad
\D C^{(p)}(t_w+\t,t_w)=\sum_{k,l}G^{(p)}_{kl}(\t)
		  \left[p_l(t_w)-p^{eq}_l\right]
\ee
and
\be\label{xi.p}
\xi^{(p)}(t_w+\t,t_w)=\L(p){\partial\over\partial t_w}
\left(\D C^{(p)}(t_w+\t,t_w)\right)
\ee
In these expressions the $p_k(t_w)=\sum_lG_{kl}(t_w)p_l^0$.
This is easy to understand from the fact that according to the chosen initial
conditions, eq.(\ref{p0.eps.a}), the variable $a$ was in equilibrium in the
beginning, $t_w\!=\!0$.
Furthermore, in case of reorientational motions I have omitted a trivial factor
${1\over 3}\mu^2$.
Note that for dipole reorientations, one is concerned with $C^{(1)}(t,t_w)$
and $\xi^{(1)}(t,t_w)$, i.e. one always has to use $L\!=\!1$ in the present
context.

If in case of $a\!=\!\Om$ it is asssumed that the reorientations proceed via
random jumps, one has to replace the factor $P_L(\cos{(\Theta)})$ in
eq.(\ref{Gl.Gq.ME}) by $\d_{L,0}$, meaning that for $L\!>\!0$ only
the diagonal element is retained.

In general, the behavior of the function $\xi^{(p)}(t,t_w)$ and all other
quantities is obtained from a numerical solution of the equations for the
Greens functions, eq.(\ref{Gl.Gq.ME})\cite{dieze97,DSHB98}.
For this purpose, the transition rates $\k_{kl}$ have to be chosen in a
prescribed way.
It is however, possible to give some general results valid in specific
situations. Consider the case of rotational random jumps, for which one has
$\L(p\!>\!0)\!=\!0$. Similarly, if the incoherent scattering function is
observed for large wave-vectors, $Q\!\gg\!1$ (or $q\!\gg\!(\d\!R)^{-1})$, one
has $j_0(Q)\!\simeq\!0$.
In both situations it is clear that $\xi^{(p)}(t,t_w)$ vanishes.
In the other extreme, namely very small wave-vectors or rotational diffusion,
one has $\L(p)\!=\!1-\eta_p$ with $\eta_Q\!=\!Q^2/6$ and
$\eta_L\!=\!L(L+1)(\Theta/2)^2$.
In both of these limiting situations, one can treat the problem in perturbation
theory, as outlined below.
\subsubsection*{${\bf \L(p)\!\to\!0}$:}
In this limit, the general expression for the Greens function,
eq.(\ref{Gp.ME}), approximately reads as
${\dot G^{(p)}}_{kl}(t)= - \k_{k}G^{(p)}_{kl}(t)$ and therefore one has
($\k_k\!:=\!\sum_{l\neq k}\k_{lk}$)
\be\label{Gp.p0}
G^{(p)}_{kl}(t)\simeq\d_{kl}e^{-\k_kt}
\ee
Using $\D p_k(t_w)\!=\!p_k(t_w)-p^{eq}_k$, one easily finds from
eq.(\ref{C.p}):
\be\label{All.p0}
C^{(p)}_{eq}(\t)\simeq\sum_ke^{-\k_k\t}p^{eq}_k
\quad\rm{and}\quad
\D C^{(p)}(t_w+\t,t_w)\simeq\sum_ke^{-\k_k\t}\D p_k(t_w)
\ee
From eq.(\ref{xi.p}) it is obvious that $\xi^{(p)}(t,t_w)\!\to\!0$ for
$\L(p)\!\to\!0$.
\subsubsection*{${\bf \L(p)\!=\!1-\eta_p}$, ${\bf\eta_p\!\ll\!1}$:}
In this case one is in the diffusive limit and one has $\eta_Q\!=\!Q^2/6$ and
$\eta_L\!=\!L(L+1)(\Theta/2)^2$, as already noted above.
Here, one can use the ME for $p\!=\!0$ as the unperturbed problem and
consider perturbation theory with a perturbation given by matrix elements
$V_{kl}\!=\!-\eta_p\k_{kl}$. This can be seen directly from the ME,
eq.(\ref{Gp.ME}) and using $\L(p)\!=\!1-\eta_p$ therein, cf. ref.\cite{DSHB98}.
The unperturbed Greens functions thus coincide with the $G_{kl}(t)$ and up to
linear order in $\eta_p$ one has:
\be\label{Gp.eta}
G^{(p)}_{kl}(t)\simeq G_{kl}(t)
-\eta_p\sum_{m,n}\int_0^t\!\!dsG_{km}(t-s)\k_{mn}G_{nl}(s)
\ee
A straightforward calculation yields:
\[
C^{(p)}(t_w+\t,t_w)\simeq 1-\eta_p\left(\lg\k\rg\t
	 +\int_0^\t\!ds\sum_k\k_k\left[p_k(s+t_w)-p^{eq}_k\right]\right)
\]
where the mean relaxation rate is defined by
\[
\lg\k\rg=\sum_k\k_kp^{eq}_k
\]
Therefore, up to linear order in $\eta_p$ one finds
\be\label{Ceq.p.eta}
C^{(p)}_{eq}(\t)\simeq e^{-\eta_p\lg\k\rg\t}
\ee
and
\be\label{DC.p.eta}
\D C^{(p)}(t_w+\t,t_w)\simeq -\eta_p\int_0^\t\!ds
	 \sum_k\k_k\left[p_k(s+t_w)-p^{eq}_k\right]e^{-\eta_p\lg\k\rg\t}
\ee

Therefore, in the limit of small wave-vectors the model predicts an
exponentially decaying intermediate incoherent scattering function,
$C^{(Q)}_{eq}(\t)\!\equiv\!S(q,\t)\!\simeq\!\exp{(-q^2D_T\t)}$, with an
apparent diffusion coefficient $D_T\!=\!(\d\!R)^2\lg\k\rg/6$\cite{DSHB98}.
In case of the rotational diffusion of molecules one accordingly has
$C^{(L)}_{eq}(\t)\!\simeq\!\exp{(-L(L+1)D_R\t)}$ with
$D_R\!=\!(\Theta/2)^2\lg\k\rg$. Remember, that in the present discussion only
$L\!=\!1$ is relevant.
Again, $\xi(t,t_w)$ is given by eq.(\ref{xi.p}).
\subsubsection*{Two-state model}
In order to illustrate the results obtained above for the energy landscape
model, I now consider only two states.
This is sufficient in order to discuss the qualitative features of the general
model. The elements of the Greens function matrix are explicitly given by
$G_{11}(t)\!=\!p^{eq}_1+p^{eq}_2e^{-\bar\k t}$,
$G_{22}(t)\!=\!p^{eq}_2+p^{eq}_1e^{-\bar\k t}$,
$G_{12}(t)\!=\!p^{eq}_1(1-e^{-\bar\k t})$ and
$G_{21}(t)\!=\!p^{eq}_2(1-e^{-\bar\k t})$. Here, I defined
$\bar\k\!=\!\k_{12}+\k_{21}$.
Furthermore, one has
$\D p_1(t_w)\!=\!\left[p^0_1p^{eq}_2-p^0_2p^{eq}_1\right]e^{-\bar\k t_w}$
and $\D p_2(t_w)\!=\!-\D p_1(t_w)$.
With these quantities one finds for the two limiting situations discussed
above:
\Be\label{TS.p0}
\hspace{-5cm}
\L(p)\to 0:\hspace{2cm}
C^{(p)}_{eq}(\t)\simeq&&\hspace{-0.6cm}
e^{-\k_{21}\t}p^{eq}_1+e^{-\k_{12}\t}p^{eq}_2\nonumber\\
\D C^{(p)}(t_w+\t,t_w)\simeq&&\hspace{-0.6cm}
\left[p^0_1p^{eq}_2-p^0_2p^{eq}_1\right]
\left[e^{-\k_{21}\t}-e^{-\k_{12}\t}\right]e^{-\bar\k t_w}
\Ee
One therefore finds that $\xi^{(p)}$ vanishes in the limit of small and large
measuring times $\t$ as well as for long waiting times $t_w$.
In addition $\xi^{(p)}$ has a maximum as a function of $\t$ for
$\t_{max}\!=\!\ln{(\k_{12}/\k_{21})}/(\k_{12}-\k_{21})$.
\Be\label{TS.p.eta}
\hspace{-2cm}
\L(p)=1-\eta_p:\hspace{2cm}&&\nonumber\\
C^{(p)}_{eq}(\t)\simeq&&\hspace{-0.6cm}e^{-\eta_p\lg\k\rg\t}\\
\D C^{(p)}(t_w+\t,t_w)\simeq&&\hspace{-0.6cm}
\eta_p\left({\k_{12}-\k_{21}\over\bar\k}\right)
\left[p^0_1p^{eq}_2-p^0_2p^{eq}_1\right]
\left[1-e^{-\bar\k\t}\right]e^{-\eta_p\lg\k\rg\t}e^{-\bar\k t_w}\nonumber
\Ee
Here, the situation regarding $\xi^{(p)}$ is analogous, the difference being
that the maximum of $\xi^{(p)}(t_w+\t,t_w)$ now occurs at
$\t_{max}\!=\!\bar \k^{-1}\ln{\left(1+{\bar \k\over\eta_p\lg\k\rg}\right)}$.
\section*{Appendix B: Environmental fluctuation model}
\setcounter{equation}{0}
\renewcommand{\theequation}{B.\arabic{equation}}
Such models are defined by the transition rates given in eq.(\ref{CMP.Om}),
where now the on-site reorientations are modelled via finite $W_k(\Om|\Om')$.
For simplicity, it is assumed in the following, that the reorientations proceed
via random rotational jumps,
$W_k(\Om|\Om')\!=\!-\G_k\d(\Om-\Om')+\G_k/(8\pi^2)$.
In addition, it has to be quantified what happens to the molecular orientation
in case of $k\!\to\!l$ transition. As in the original model of Beckert and
Pfeifer\cite{BP65}, it will be assumed that either no change at all,
$\L(\Om|\Om')\!=\!1$, or a random rotation, $\L(\Om|\Om')\!=\!1/(8\pi^2)$,
takes place.
Proceeding exactly in the same way as in Appendix A, the Greens functions are
obtained from:
\be\label{Gl.BP}
\dot{G}^{(L)}_{kl}(t)=-\left(\G_k+\k_k\right)G^{(L)}_{kl}(t)
		       +c_L\sum_n\k_{kn}G^{(L)}_{nl}(t)
\ee
Here, I defined $c_L\!=\!1$, if $\L(\Om|\Om')\!=\!1$ and
$c_L=\d_{L,0}$ if $\L(\Om|\Om')\!=\!1/(8\pi^2)$\cite{sill96}.
As in the case of the energy landscape model, the relevant quantities are
given by eqns.(\ref{C.p}) and (\ref{xi.p}) if again it is assumed that there
is no correlation initially, $p(\{\e_k,\Om\};0)\!=\!{1\over 8\pi^2}p^{eq}_k$.

In the following, the explicit expressions will be given for the simple
two-state model discussed in Sect.IV.
Here, I consider the case $c_L\!=\!1$.
As in the text, the states are denoted by $f$ and $s$.
The equilibrium populations of the states are given by $p_s^{eq}$ and
$p_f^{eq}$, the reorientation rates are $\G_s$, $\G_f$ and the exchange rates
are denoted by $\k_{sf}$ and $\k_{fs}$.
The solution of eq.(\ref{Gl.BP}) poses no problem and one finds, again
omitting a factor $\mu^2/3$:
\Be\label{C.BP}
C^{(1)}_{eq}(\t)=&&\hspace{-0.6cm}
     {1\over 2w}\left[z_fe^{-\L_f\t}+z_se^{-\L_s\t}\right]\\
\D C^{(1)}(t_w+\t,t_w)=&&\hspace{-0.6cm}
     {\g\over w}\left[p_f^0p_s^{eq}-p_s^0p_f^{eq}\right]
     \left[e^{-\L_f\t}-e^{-\L_s\t}\right]e^{-2\bar{\k} t_w}\nonumber
\Ee
Using the abbreviations $\bar{\k}\!=\!(1/2)(\k_{sf}+\k_{fs})$,
$\bar{\G}\!=\!(1/2)(\G_f+\G_s)$, $\g\!=\!(1/2)(\G_f-\G_s)$ and
$\D_\k\!=\!(1/2)(\k_{fs}-\k_{sf})$ one has for the effective decay rates
$\L_{f,s}=\bar{\G}+\bar{\k}\pm\sqrt{\left(\D_\k+\g\right)^2+\k_{fs}\k_{sf}}$.
Furthermore, I used $w=\sqrt{\g^2+2\g\D_\k+\bar{\k}^2}$ and
$z_{f/s}\!=\!w\mp\bar{\k}\pm\g(p_f^{eq}-p_s^{eq})$.
Finally, for the function $\xi$ one finds:
\be\label{xi.BP}
\xi^{(1)}(t,t_w)=-(2\bar{\k})\D C^{(1)}(t,t_w)
\ee

The most important thing that changes, if one assumes that a random
reorientation is associated with every $k\!\to\!l$ transition,
$c_L\!=\!\d_{L,0}$, in eq.(\ref{Gl.BP}) is that one now finds:
\be\label{xi.BP.RJ}
\xi^{(1)}(t,t_w)=0\quad\mbox{for}\quad c_L\!=\!\d_{L,0}
\ee
\end{appendix}

\section*{Figure captions}
\begin{description}
\item[Fig.1 : ]
$\b^{-1}\chi(t_w+\t,t_w)$ versus $C(t_w+\t,t_w)$, for $x\!=\!0.3$ and
for various waiting times $t_w$.
The dotted lines represent the slopes expected for ordinary FDT (slope: $-1$)
and an effective temperature of $T_{\rm eff}=T_c$ (slope: $-x$).
\item[Fig.2 : ]
$[-\xi(t_w+\t,t_w)]$ versus $\t$, for $x\!=\!0.3$ for various waiting times
$t_w$. Upper panel: linear scale, lower panel: logarithmic scale.
In the lower panel, the approximate expression, eq.(\ref{Ksi.rcm.ap}), is
shown as thin dotted lines.
\end{description}
\newpage
\begin{figure}
\includegraphics[width=16cm]{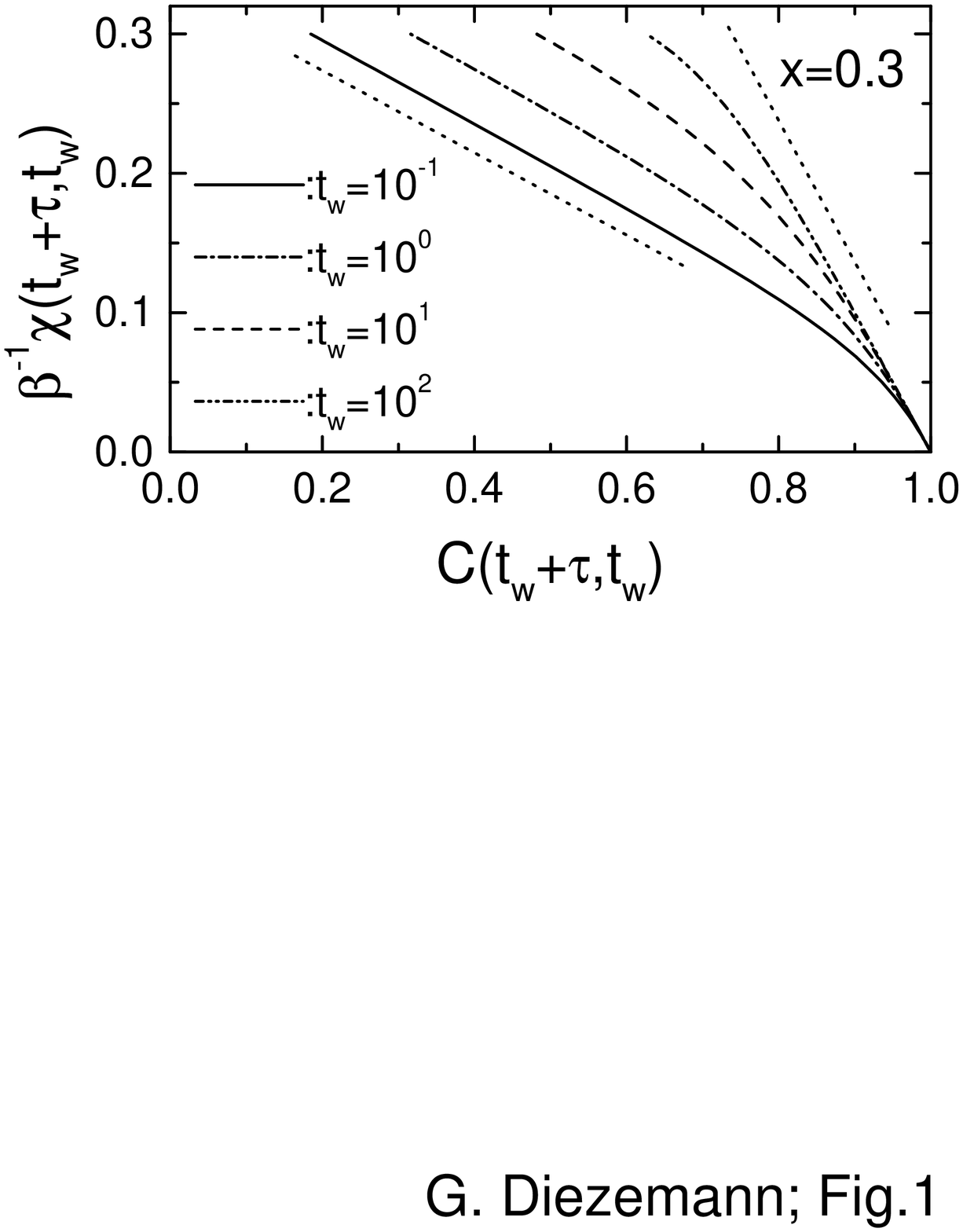}
\end{figure}
\newpage
\begin{figure}
\includegraphics[width=16cm]{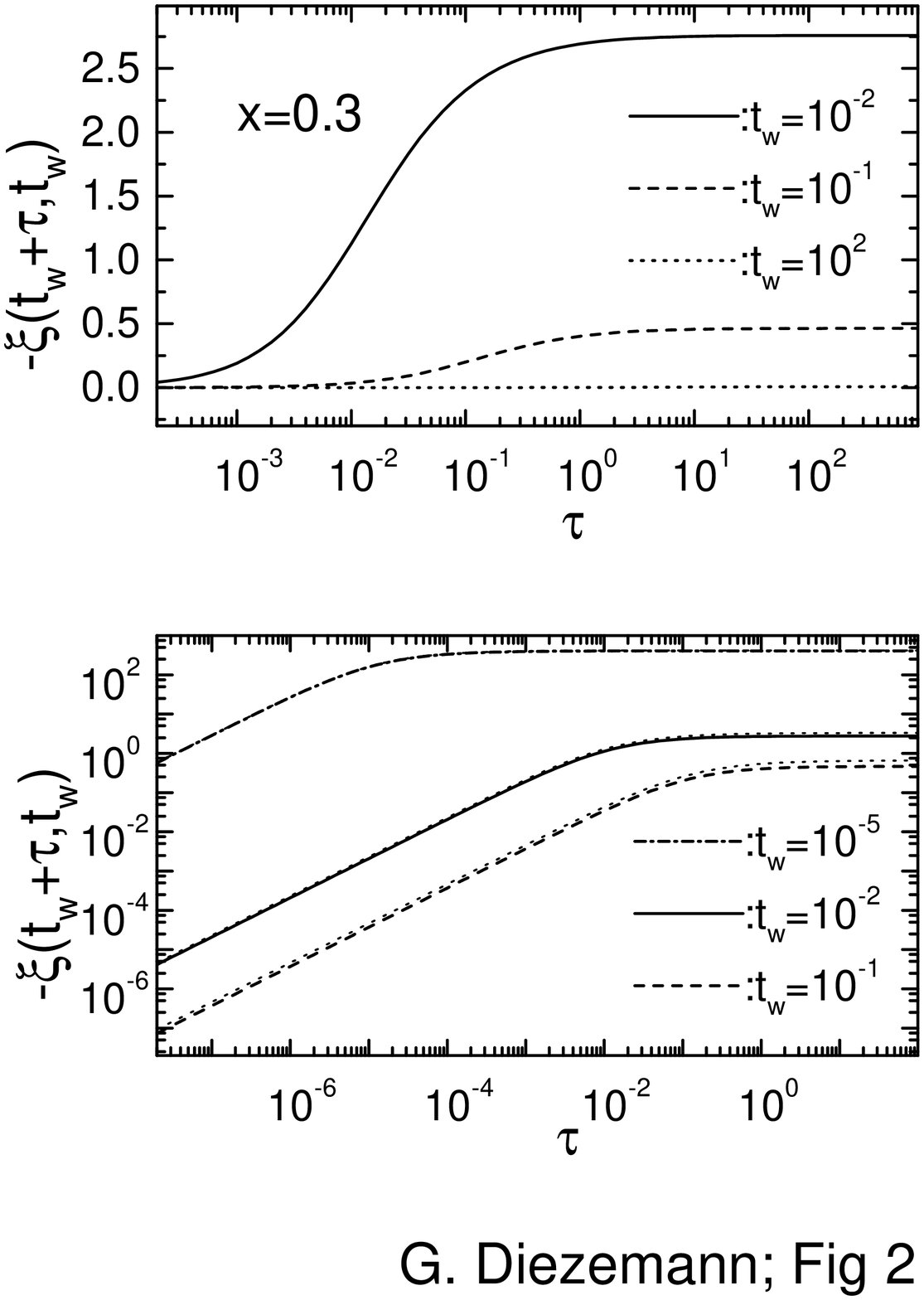}
\end{figure}
\end{document}